# Spatially Multiplexed Interferometric Microscopy with one-dimensional diffraction grating

José Ángel Picazo-Bueno[1], Maciej Trusiak[2], Javier García[1], and Vicente Micó[1]

[1] Departamento de Óptica y Optometría y Ciencias de la Visión, Facultad de Física, Universitat de Valencia, c/ Doctor Moliner 50, Burjassot 46100, Spain
[2] Warsaw University of Technology, Institute of Micromechanics and Photonics, 8 Sw. A. Bodoli St., 02-525 Warsaw, Poland

E-mail: j.angel.picazo@uv.es

**Abstract**

Digital holographic microscopy (DHM) applied to quantitative phase imaging (QPI) has been successfully demonstrated as a powerful label-free method to analyse the optical properties of cells. Spatially multiplexed interferometric microscopy (SMIM) is a DHM technique that implements a common-path interferometric layout in the embodiment of a standard microscope to achieve QPI. More concretely, SMIM introduces three minimal modifications: 1) replaces the broadband illumination of the microscope by a coherent or partially coherent light source, 2) divides the input plane into two or three regions for transmission in parallel of both imaging and reference beams, and 3) includes a one-dimensional diffraction grating or a beam splitter cube for holographic recording. Hence, SMIM is a cost-effective, extremely simple, and highly stable manner of converting a standard bright field microscope into a holographic one. The goal of this contribution is to provide a review of the SMIM approaches implemented using a one-dimensional (1D) diffraction grating, and highlight vast range of capabilities including super-resolved, reflective, transflective, noise-reduced and single-shot slightly off-axis amplitude and phase imaging.

## 1. Introduction

Nowadays, quantitative phase imaging (QPI) has become a powerful, label-free imaging approach that permits accurate analysis of structure and dynamics of micrometric specimens, such as cells and tissues, very useful for medical diagnosis [1–5]. QPI provides quantitative information about both local thickness and refractive index (RI) distributions of the samples with high accuracy [6–12]. QPI is usually achieved by digital holographic microscopy (DHM) [13–17]. DHM combines all advantages of classical holography [18, 19], optical microscopy [20], and numerical computations [21] in terms of high-quality imaging, whole-object wavefront recovery, and numerical processing capabilities [6, 8, 17, 22, 23]. At the same time, it avoids issues of conventional microscopy such as the limited depth of focus in high NA objectives [23]. In addition, DHM enables visualization of phase samples using a label-free, non-scanning, real-time, non-contact, and static (no mechanical movement) operation principle [6, 8, 22, 24]. Recently, DHM has been successfully employed in many relevant applications in cell biology such as three-dimensional (3D) analysis and monitoring of cell migration [25–32], automated cell counting, recognition, and classification [33–41], and manipulating and testing biomechanical properties of cells [42–44]. For those reasons, DHM emerged as a powerful and versatile technique applied to many fields of knowledge, such as Biophotonics, Life Sciences and Medicine [3–5, 16, 45].

The interferometric principle of DHM makes imperative the use of interferometric configurations such as Mach-Zehnder [46, 47], Michelson [48], Linnik [49] and common-path interferometers [50, 51], just to cite a few. Among them, Mach-Zehnder arrangement is the main type of interferometer used for transmissive objects [6, 8, 16, 17, 46, 47], whereas Michelson architecture is the most assembled in reflective configuration [14, 48, 52–55]. However, since in common-path interferometers (CPIs) both interferometric beams follow closely the same optical path and pass through the same microscope lens, they present substantial advantages with respect to Mach-Zehnder and Michelson configurations in terms of robustness, simplicity, and stability. Additionally, CPI demands less optical elements, can be assembled in a more compact way, and is relatively insensitive to vibrations [56–61].

Basically, CPIs can be classified into three categories according to the way of generating the reference beam [62]. When such a beam is synthesized from the imaging beam after passing through the microscope lens, a reference-generation (R-G) CPI is defined [50, 51, 63–65]. R-G CPIs preserve whole field of view (FOV), but a rather complex opto-mechanical stage must be set up at the image space instead. Another type of CPIs take advantage of clear regions in the surrounding areas of sparse samples for reference beam generation, being referred as self-interference (S-I) CPIs [53, 60, 61, 66]. There, S-I CPIs involve just a few optical elements to make the imaging beam interfere with a shifted version of itself. S-I CPI cannot be used with dense samples, which is clearly a shortcoming. The third group of CPIs can be named as spatially-multiplexed (S-M) CPIs. S-M CPIs get the reference beam transmission by clearing an appropriate area at one side of the object region at the input plane, and use minimal elements for interferometric recording [56, 57, 67–70]. Such a S-M FOV permits the inspection of not only sparse but also dense samples at the expense of reducing the useful FOV.

In recent years, many researchers have been focused on improving DHM capabilities in terms of robustness, simplicity, usability, accuracy, and price [45, 53, 60, 61, 64–66, 71–76]. Thus, equipping regular microscopes with coherence sensing capabilities has become a very appealing option for that purpose [8, 50, 53, 74, 75]. In line with that, different approaches were reported such as diffraction phase microscopy (DPM) [50], quadriwave lateral shearing interferometry (QLSI) [74], and Michelson interferometer layout (MIL) [53], just to cite a few. Briefly, DPM implements a R-G CPI layout in which one of the split imaging beams is spatially filtered at the Fourier plane by using a pinhole for reference beam generation, and then both beams interfere at the recording plane. By contrast, a S-I CPI is assembled in MIL approach using a Michelson architecture, so it is limited to sparse samples. Finally, QLSI utilizes a shearing interferometer based on a modified Hartmann mask to sense wavefront derivatives and digitally integrate the wavefront itself. Closely related to those techniques, recently appeared a novel technique named spatially multiplexed interferometric microscopy (SMIM) that updates a standard microscope with holographic capabilities in a stable, simple, and low-cost way. SMIM has the advantages of not needing to include any pinhole mask to generate the reference beam when comparing with DPM, and additionally is not restricted only to sparse samples unlike MIL, since SMIM implements a S-M CPI configuration in the embodiment of the microscope by leaving a clear region in the FOV.

This review collects all contributions related to SMIM technique implemented using a one-dimensional (1D) diffraction grating for holographic recording. The paper is organized as follows. Section 2 describes the working principle of SMIM technique. Section 3 presents all SMIM experimental validations. Finally, Section 4 concludes the review.

## 2. Description of SMIM using a diffraction grating

The SMIM concept was first validated at the laboratory for superresolution (SR) purposes [56, 57, 77], and then implemented in a regular microscope [62, 78–83], giving rise to SMIM itself. SMIM is based on the idea of implementing a S-M CPI architecture in the embodiment of a conventional bright field microscope to convert it into a holographic one. For that purpose, SMIM introduces the following three minimal modifications: 1) the replacement of the broadband illumination by a (partially) coherent light source, 2) the definition of an input plane with a specific spatial distribution for tranmission in parallel of both reference and imaging beams, and 3) the insertion of a 1D diffraction grating to provide an interference pattern at the recording plane. Figure 1 shows the optical scheme and ray tracing of a regular microscope before [see figure 1(a)] and after SMIM implementation [see figure 1(b)], where the main elements and planes of the microscope are pointed in figure 1(a), whereas the main changes introduced by SMIM are highlighted in figure 1(b).

Regarding illumination requirements, SMIM replaces the broadband microscope light source by a coherent (or partially coherent) spherical divergent illumination with enough coherence to produce interference patterns at the sensor plane. On the other hand, SMIM divides the input plane into two or three regions. Figure 1 depicts the situation in which the input plane is spatially multiplexed into three regions. Thus, the input plane includes an object (O), reference (R) and blocking

(X) region with the same size [see figure 1(b)]. The O region contains the sample to be inspected and is centered with respect to the optical axis of the microscope, whereas the R and X regions are a clear (no sample) and blocked (no light) region, respectively. Those regions are located at both sides of the O region, and serve to allow reference beam transmission (R region) and to avoid spurious interferences (X region). As consequence, the useful FOV is reduced to 1/3 of the available one due to such spatial multiplexing. Nonetheless, the recording area of a digital sensor is usually smaller than the image of the FOV provided by the microscope. In such a case, it is possible to make that only the image of the O region (O') falls into the sensitive area of the digital sensor by properly selecting its dimensions and/or the magnification of the optical system, and the R' and X' regions (images of R and X regions, respectively) fall outside of such a region.

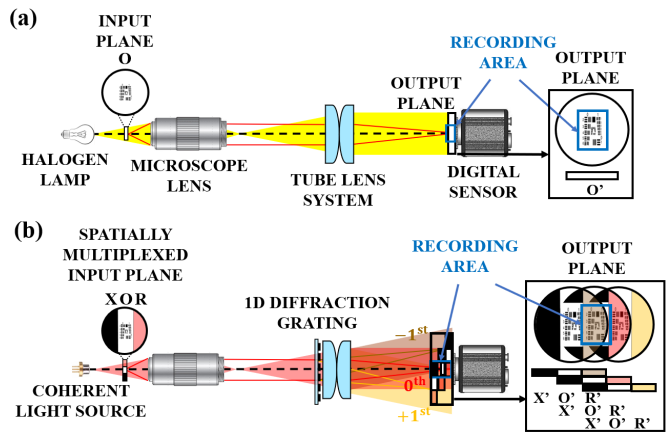

**Figure 1.** Optical scheme of a regular microscope before (a) and after (b) SMIM application, pointing the modifications introduced by SMIM.

Finally, a 1D diffraction grating is properly placed to produce an interference pattern between the O', R' and X' regions onto the recording area, thus creating a digital hologram. Assuming a 1D sinusoidal grating, three shifted replicas of the image of the FOV are generated at the output plane [see figure 1(b)], corresponding to the $0^{th}$, $+1^{st}$, and $-1^{st}$ diffraction orders. Then, an interference pattern between R', O' and X' regions can be recorded when the lateral displacement between replicas, at the output plane, is equal to 1/3 of the image of whole FOV, as it is shown in figure 1(b). That can be achieved by a proper selection of the spatial frequency of the grating. Since the X' region of the $+1^{st}$ order does not contain light (blocked at the input plane), the digital sensor will record a digital hologram created by the coherent overlapping of the R' and O' regions coming from $-1^{st}$ and $0^{th}$ orders, respectively. On the other hand, the axial position of the grating with respect to the Fourier plane of the microscope plays an important role in both the grating frequency to be selected and the holographic configuration. Two main scenarios were carefully studied in [56, 57, 77]. When the grating is placed far from the Fourier plane, an off-axis hologram can be recorded with a big enough grating frequency. By contrast, if the grating is located close or in the Fourier plane, both R' and O' beams interfere on-axis since they propagate with the same propagation angle. SMIM approaches are more closely related to the first scenario since the diffraction grating is usually placed just before the tube lens system of the microscope.

## 3. Experimental validations

So far, all reported SMIM approaches were implemented in the embodiment of an infinity-corrected BX60 Olympus upright microscope [62, 78–83]. Furthermore, in the majority of SMIM layouts, a Ronchi grating was employed as interferometric element [62, 78–81]. In all those cases, the grating was inserted in the analyser insertion slot of the microscope, just before the tube lens system. The use of a Ronchi grating instead of a sinusoidal one did not introduce any additional issue regarding the presence of high diffraction orders, since the diffraction efficiency of even orders is zero and odd orders fall outside the recording area because of their high diffraction angles. In addition, the location of the diffraction grating in the infinity space of the microscope avoided both the introduction of aberrations and restrictions due to field or aperture diaphragm effects. On the other hand, the microscope halogen lamp was turned off and a laser diode was externally inserted just below the XYZ translation stage of the microscope for coherent illumination.

Along different contributions, SMIM arrangements were experimentally validated involving different technical objects, such as resolution test targets or microbeads, as well as several types of biological samples like red blood cells (RBCs), swine sperm cells (SSCs), and prostate cancer cells (PCCs), for different microscope lenses. In those cases, a specific chamber was not designed for application, but the clear regions present in the surroundings areas of the samples were considered for reference beam transmission.

### 3.1 Conventional SMIM

The first SMIM implementation was carried out by Micó et al. [62]. In that contribution, the authors used a DVD laser diode source (650 nm) for spherical divergent coherent illumination, defined an input plane spatially multiplexed into three regions (as described in Section 2), and inserted a Ronchi grating of 80 lp/mm spatial frequency. For digital recording, they employed a charge-coupled device (CCD) sensor (Basler A312f, 582x 782 pixels, 8.3 μm pixel size, 12 bits/pixel). In addition, the approach was experimentally validated for three infinity corrected microscope lenses: Olympus UMPlanFl 5X/0.15NA, 10X/0.3NA, and 20X/0.46NA objectives. Finally, the magnification of the imaging system, the characteristics of the digital sensor and the high frequency of the grating provided off-axis holograms and, therefore, single-shot capability by means of the application of Fourier filtering methods for phase retrieval.

There, an USAF resolution test target served as a sample for calibrating the conventional SMIM approach. The region containing the maximum resolution elements was considered as object region (O region). On the other hand, the largest clear region at one side of the O region was used as reference region (R region), whereas the other side was manually blocked (X region). Figure 2 presents the experimental results achieved for 5X, 10X, and 20X microscope lenses, including the recorded interferograms [figure 2(a1)-(c1)] with the area enclosed in the white rectangles magnified to clearly show the fringes, their Fourier transforms, showing the separation between orders and the apertures defined for filtering [figure 2(a2)-(c2)], and the images obtained after Fourier filtering application in figure 2(a3)-(c3).

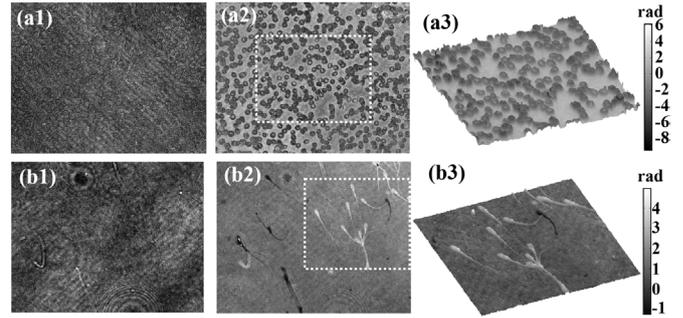

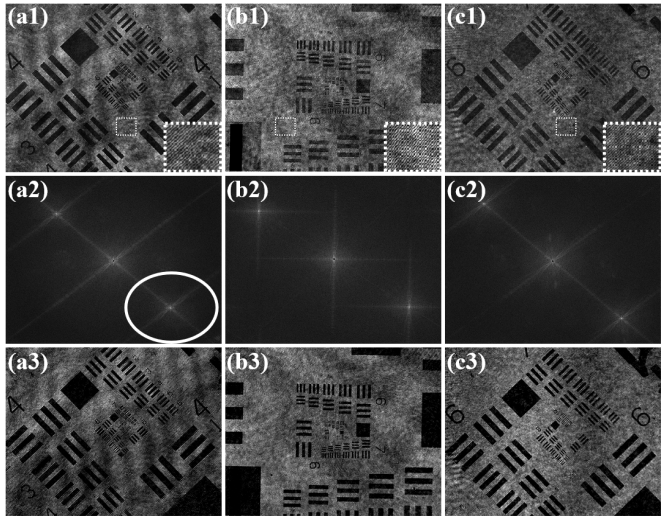

**Figure 2.** Experimental validation of conventional SMIM involving an USAF resolution test target. Columns (a)-(c): results for the 5X, 10X, and 20X lenses, respectively. Rows (1)-(3): holograms, Fourier transforms of the holograms, and retrieved intensity images, respectively. (Images adapted from [62] by permission of OSA).

Then, conventional SMIM approach was validated using two types of static biological samples (SSCs and human RBCs) for the 10X and 20X microscope lenses. In those cases, both biosamples were arranged to produce an abrupt change between the O and R region for proper SMIM performance. Figure 3 depicts the holograms [figure 3(a1) and (b1)], and the wrapped [figure 3(a2) and (b2)] and 3D unwrapped [figure 3(a3) and (b3)] phase distributions for both biosamples using the 20X/0.46NA objective. In addition, quantitative phase values provided by SMIM were successfully compared with the obtained by a conventional DHM layout (Mach-Zehnder architecture) for the case of SSCs and the 20X lens, thus validating conventional SMIM for QPI.

**Figure 3.** Experimental results of conventional SMIM for the cases of RBCs and SSCs using a 20X/0.46NA objective. Column (1) the holograms, column (2) the retrieved phase images, and column (3) the perspective view of unwrapped phase maps of the regions enclosed in the white rectangles in column (2), respectively. (Images derived from Ref. [62] by permission of OSA).

### 3.2 Superresolved SMIM

The main drawback of SMIM technique is the necessity to leave a clear region of the FOV for reference beam transmission, thus significantly reducing the useful FOV to one third (or one half, at best) of the available one. Such a reduction may be critical when implementing SMIM with high NA objectives. On the other hand, lower NA microscope lenses, which typically provide lower magnifications and spatial resolutions but larger FOVs than the higher ones, are also penalized in FOV with the application of SMIM. Nevertheless, this fact can be somehow compensated with a resolution enhancement, so that the images provided by SMIM using a low NA objective can present similar features, regarding spatial resolution and FOV, that the achieved ones by a higher NA lens.

In order to overcome such an issue, SMIM was upgraded with superresolution (SR) capability with the technique named superresolved spatially multiplexed interferometric microscopy (S2MIM) [78]. S2MIM employs an angular and time multiplexing approach to achieve a resolution gain factor of 2. Briefly, S2MIM is sequentially implemented by using several tilted beam illuminations achieved by lateral displacements of the coherent source to a set of four off-axis positions [see figure 4(a)]. Then, the complementary spatial-frequency content provided by each tilted illumination is transmitted on-axis through the imaging aperture, and recorded and retrieved by holographic procedures [see figure 4(b)]. The only restriction is that the illumination angle $\theta_{illum}$ must be below the defined by the NA of the objective. Finally, all that information is used to generate a synthetic aperture (SA) with a cutoff frequency higher than the provided by the optical system, thus allowing SR imaging by the Fourier transform of the SA.

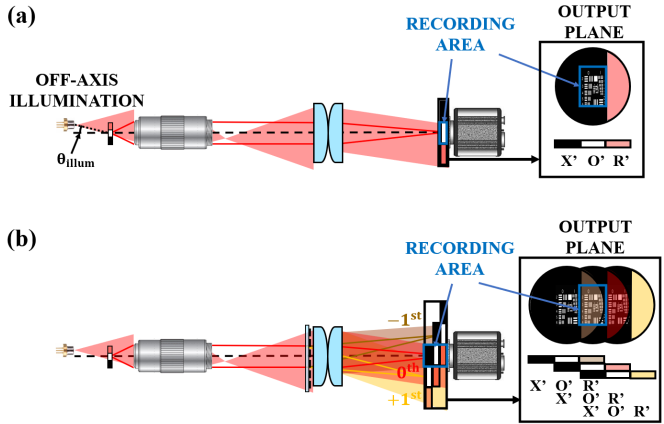

**Figure 4.** Scheme of the S2MIM layout for the case of tilted beam illumination before (a) and after (b) diffraction grating insertion.

In the experimental setup of S2MIM, a laser diode illuminated the input plane divided into three regions, and a slightly off-axis holographic configuration and further temporal phase-shifting (P-S) method were implemented to record and retrieve the different bandpasses. The experimental validation was carried out for two different low NA objectives (Olympus UMPlanFl 2X/0.05NA and 5X/0.15NA) employing different components and resolution objects: (1) a VCSEL source (1 mW optical power, 850 nm) and a DVD laser diode source (5 mW optical power, 650 nm), (2) a Ronchi grating of 80 and 50 lp/mm, and (3) two different resolution test targets (NBS 1963A and USAF test targets), respectively. To record the whole set of phase-shifted holograms, the Ronchi gratings were placed onto a motorized translation stage (Newport, model ESP300).

Figure 5 includes the results for the case of the USAF resolution test target and the 5X microscope lens. The intensity image obtained with conventional SMIM is included in figure 5(a), which corresponds to the central aperture presented in figure 5(b) with a solid white circle. Then, the coherent addition of the four additional bandpasses, depicted with dashed white circles in figure 5(b), generated a SA and the final SR image was obtained by Fourier transform of the SA [see figure 5(c)]. To clearly show the resolution improvement, a comparative plot along the vertical dashed lines placed at the black-blue insets in figure 5(a) and 5(c) is presented in figure 5(d), in which one can see how the elements of the group 8 are or not well resolved with S2MIM and SMIM, respectively. The last element resolved after S2MIM was G9-E1 corresponding with a resolution of 1.95 μm, whereas the resolution achieved by conventional SMIM was 3.91 μm (G8-E1). That means a resolution gain factor of 2, thus obtaining the same resolution limit as compared against the image provided by an Olympus UMPlanFl 10X/0.30NA under the same on-axis illumination conditions, as the authors reported in [78]. Finally, the second experimental validation performed with a 2X lens and the NBS 1963A resolution target demonstrated, once again, the resolution gain factor of 2 by using S2MIM (conventional imaging provided a resolution limit of 13.9 μm, whereas S2MIM was able to resolve the 143 lp/mm element, corresponding to a resolution of 7 μm).

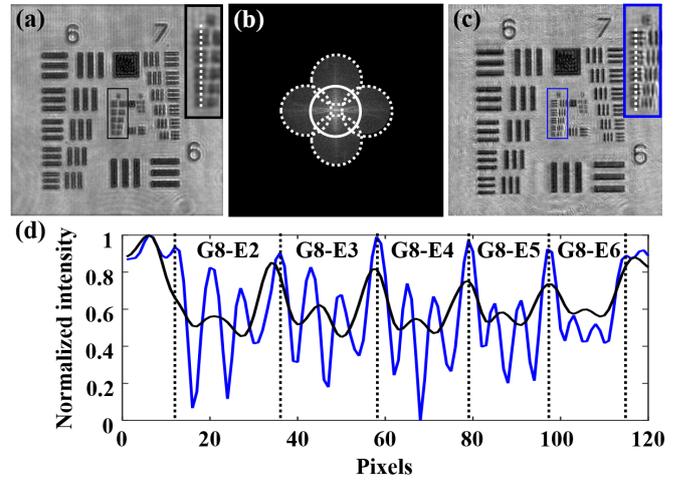

**Figure 5.** Final results of the experimental validation of S2MIM involving a 5X/0.15NA objective and an USAF resolution test target: (a) the low-resolution intensity image provided by conventional SMIM, (b) the SA coming from the addition of the four additional off-axis apertures, (c) the SR image provided by S2MIM, and (d) the plots along the vertical white lines included in (a) and (c). (Images adapted from [78] by permission of OSA).

### 3.3 Opposed-view SMIM

SMIM can be also applied to the study of opaque samples, as it is demonstrated in [84] by the implementation of SMIM working under reflection imaging mode. There, the authors additionally proposed a novel SMIM transflective modality to simultaneously record the light transmitted and reflected by a sample, named opposed-view spatially multiplexed interferometric microscopy (OV-SMIM). That allowed to generalize the SMIM concept for the study of not only transparent but also opaque and transflective objects.

Figure 6 includes a scheme of the experimental setup developed for the validation of the OV-SMIM technique. There, the sample was illuminated in transmission and reflection using the green (520 nm) and red (635 nm) illuminations of two different fiber-coupled diode lasers (OSI Laser Diode, TCW RGBS-400R and Blue Sky Research, SpectraTec 4). On the other hand, the FOV was divided into three regions, in which the R region for transmission was employed as X region in reflection, and vice versa. A reflecting surface served, not only to block the light passing through the object, but also to provide a reference beam in reflection. A Ronchi grating of 40 lp/mm frequency placed onto a motorized translation stage (Newport, model ESP300) provided slightly off-axis P-S holograms for temporal P-S application.

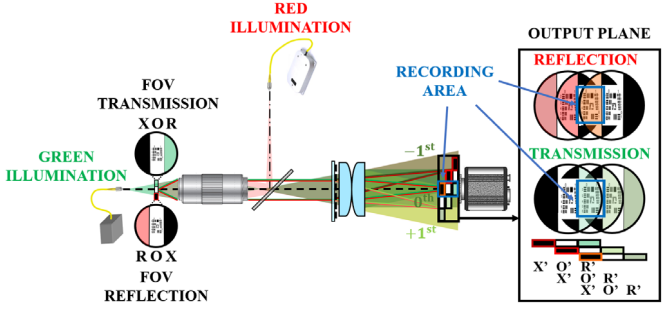

**Figure 6.** Optical scheme and ray tracing of the OV-SMIM approach.

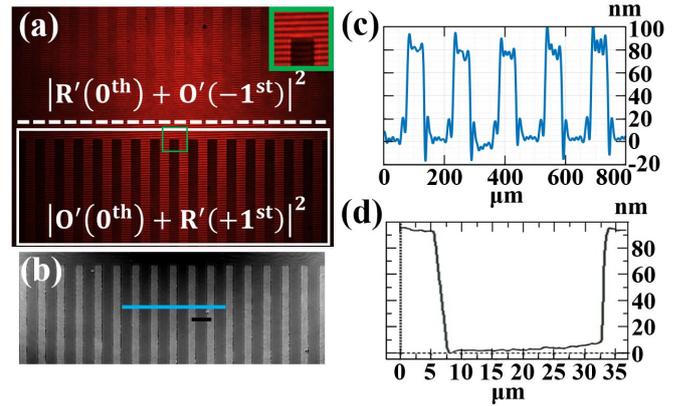

**Figure 7.** Experimental validation of *OV-SMIM* for reflective imaging mode: (a) reflective hologram, (b) retrieved phase distribution of the region enclosed in the solid white rectangle included in (a) after applying temporal P-S method, and (c) and (d) thickness profiles along blue and black lines included in (b) provided by reflective SMIM and AFM, respectively. (Images adapted from [84] by permission of OSA).

In [84], the authors validated first SMIM for reflective imaging mode. For that purpose, a Ronchi grating with a frequency of 20 lp/mm was regarded as opaque sample and imaged with an Olympus UMPlanFl 5X/0.15NA objective. In that case, the reflective coating frame of the grating served as R region, while the digital sensor only covered the R′ and O′ regions (but not X′). The results are shown in figure 7, where the hologram is depicted in figure 7(a). The hologram includes a dashed white line in order to divide the output plane intensity distribution in useful interferogram (lower half) and useless interference pattern (upper half). In addition, the region enclosed in the green square is magnified to clearly see the fringes. Then, a P-S algorithm is applied to the region enclosed in the white rectangle included in figure 7(a) to recover the phase distribution of the grating [see figure 7(b)]. Such a phase map can be easily converted to height distribution [84], and a plot along the blue line included in figure 7(b) can be performed to obtain a thickness profile of the chromium layer of the grating, as it is depicted in figure 7(c). From this profile, the authors computed an average thickness value of 82 nm using reflective SMIM. Finally, in order to validate such a value, a characterization of one grating profiles was performed by using an atomic force microscope (AFM), which provided a height profile [see figure 7(d)] along the black line included in figure 7(b). AFM measurement provided a thickness value of around 85 nm, showing high agreement between the values obtained by reflective SMIM and AFM characterization, thus validating SMIM under reflective modality.

After that, the experimental validation of the OV-SMIM approach was performed involving first, a NBS 1963A resolution test target, and second, several microbeads of 45 µm in diameter immersed into water (sparse mode). There, a wavelength multiplexing approach was followed and a direct demosaicing employed for separating the information from the transflective hologram. For the case of the test target, imaged with the same 5X/0.15NA objective, the authors demonstrated that the quantitative phase distributions achieved in transmission and reflection presented similar phase values (leaving aside differences in resolution) once the optical paths were compensated (reflected light travels twice along the same optical path) [84].

On the other hand, the microbeads were imaged with an Olympus UMPlanFl 10X/0.30NA objective to get higher imaging performance, and a mirrored surface was placed at one side of the O region acting as R/X regions for reflection/transmission. Figure 8 presents the results for the case of the spheres. One of the transflective holograms recorded for temporal P-S method is shown in figure 8(a). After demosaicing and further P-S application, the phase distributions of the light transmitted and reflected by the microspheres were achieved and, after that, the height distributions were computed by using the pertinent expressions [84]. Figures 8(b) and 8(c) depict such thickness maps for the microbeads enclosed in the white rectangle included in figure 8(a), after the wavelength dependence is removed and the optical path for transmission and reflection is equalized. The height profiles along green and red solid lines included in figure 8(b) and 8(c), respectively, and presented in figure 8(d), clearly show the concordance between such profiles, and provided a height value about 7.4 µm. However, this value was far from the diameter value given by manufactures for those microspheres (45 µm) due to the use of a modest/low NA objective for imaging. Despite that, the authors were able to make a rough estimation of the refractive index (RI) of the two beads, whose RI profile is included in figure 8(e). Finally, an average value of 1.59 was reached considering only the regions of the beads enclosed in the full width at half maximum (depicted with the grey rectangles in figure 8(d) and 8(e)), a value really close to the theoretical one for polystyrene material at 520 nm, which is 1.6.

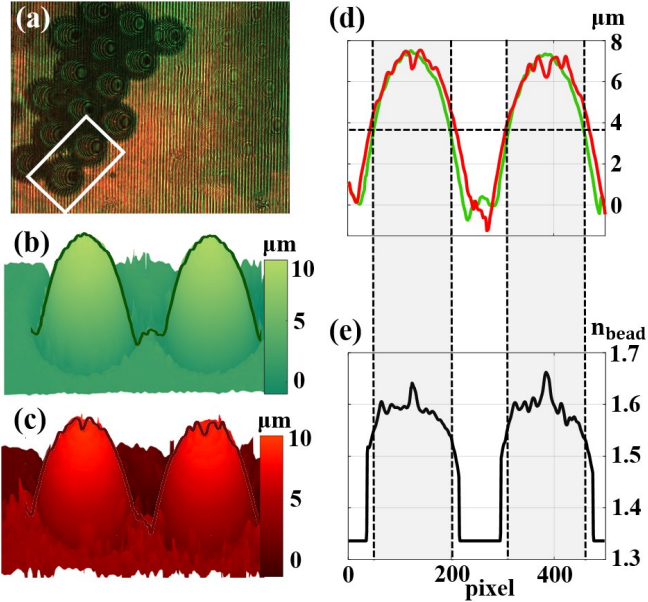

**Figure 8.** Experimental results of OV-SMIM involving 45 µm beads: (a) transflective hologram, (b) and (c) perspective views of thickness distributions obtained for the two microspheres enclosed in the white rectangle included in (a) in transmission and reflection modalities, respectively, and (d) and (e) comparative thickness and RI plots along the solid green and red lines included in (b) and (c), respectively. Shadowed rectangles in (d) and (e) indicate the microbeads region considered for averaged RI computation. (Image derived from [84] by permission of OSA).

*3.4 SMIM with partially coherent illumination*

SMIM was also implemented under partially coherent illumination to reduce the coherent noise of phase images [80, 81]. In those contributions, the light source was not a laser diode but a superluminescent diode (SLD), which provided partially coherent (not only temporal incoherence but also spatial coherence) quasi-point illumination. Partially coherent illumination was previously employed in DHM layouts [47, 52, 85–90] for the reduction of the coherent noise coming from speckle, coherent artifacts, and multiple reflections, thus increasing the quality of the retrieved phase images. For that purpose, the authors employed a SLD source from Exalos (Model EXS6501-B001, 10 mW optical power, $\lambda = 650$ nm central wavelength, $\Delta\lambda = 6$ nm spectral bandwidth). The coherence length [91] of the SLD was

$$L_C = k \frac{\lambda^2}{\Delta\lambda} \sim 50 \text{ µm}$$

being k = 0.66 for Gaussian spectrum. Such low coherence length forced to use a grating with lower spatial frequency (20 lp/mm) for interferometric recording. That prevented an off-axis holographic recording and further Fourier filtering application for real-time analysis. Instead, a temporal P-S approach was followed in [80], whereas a Hilbert-Huang algorithm was applied in [81] for phase retrieval.

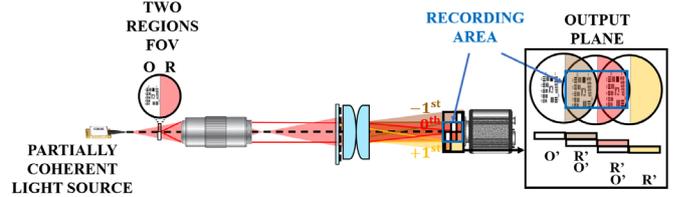

**Figure 9.** Scheme of SMIM approach with partially coherent illumination, where the main changes regarding illumination and FOV spatial multiplexing are identified.

The other serious constraint of the previous SMIM approaches was the division of the FOV into three regions [62, 78, 79]. That significantly reduced the useful FOV to just one third of the available one. However, the blocking region X included in those contributions can be avoided by a slightly different spatial distribution at the input and output planes [62, 80, 81]. Figure 9 shows the optical scheme of the SMIM layout implemented in Refs. [80, 81] without the requirement of such a blocking area. In those cases, the FOV was spatially multiplexed into two equal regions (instead of three): an object (O) and a reference (R) region, thus increasing the useful FOV from one third to one half of the available one. That was accomplished by having the O and R regions symmetric with respect to the optical axis rather than the O region centered with it. Nevertheless, the main drawback was: either the digital sensor had to be laterally shifted and placed in an off-axis position (as proposed in [62]), therefore modifying the exit port of the microscope, or half of the sensitive area of the CCD had to be wasted (as implemented in [80, 81]) in order to have the CCD centered with the optical axis of the microscope.

Summarizing, in Refs. [80, 81], a partially coherent light source illuminated the input plane, which presented a division into two equal regions (O and R), and a low frequency grating was employed to achieved an interference pattern of the regions O and R coming from the $0^{th}$ and $+1^{st}$ diffraction orders, respectively, covering one half of the CCD and losing the other half (see figure 9).

In Ref. [80], the experimental validation started with a calibration stage involving microbeads (standard monodisperse polystyrene microspheres of 45 µm mean diameter in aqueous suspension) and an USAF resolution test target. Those synthetic samples were imaged with the Olympus UMPlanFl 10X/0.3NA and 20X/0.46NA objectives, respectively. The sparsity of the beads assured the presence of clear regions in the surrounding areas of the microspheres used for reference beam transmission, whereas the clear region between 4 and 6 groups was regarded as R region. A motorized linear translation stage (Newport, model ESP300) was employed for a precise displacement of the grating (grating motion step of 2.5 µm) and a set of 40 P-S holograms (2 full P-S cycles) was recorded and processed for amplitude and phase retrieval.

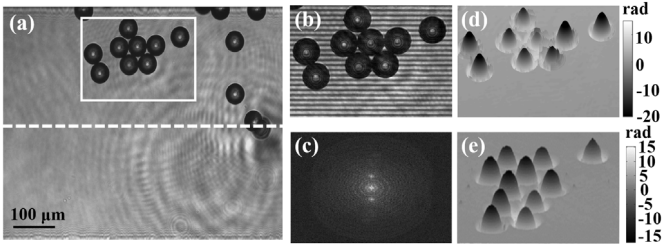

**Figure 10.** Experimental results of SMIM approach with partially coherent illumination [80] for the case of 45 μm spheres: (a) Intensity image of whole FOV showing the O (upper half) and R (lower half) regions defined by the spatial multiplexing, (b) ROI of one of the P-S holograms recorded for P-S application and (c) its Fourier transform, and (d) and (e) perspective views of the phase distributions retrieved by the SMIM approach and by a conventional DHM layout. (Images derived from [80] with SPIE permission).

Figure 10 shows the results obtained for the case of the microbeads. The direct intensity image recorded before inserting the grating, in which we can clearly appreciate the spatial multiplexing of the FOV into the O (upper part) and the R (lower part) regions is included in figure 10(a). In addition, a region of interest (ROI) of a hologram captured during the P-S recording process is presented in figure 10(b), and its Fourier transform is depicted in figure 10(c), showing the overlapping between orders. After P-S implementation, the phase distribution of the microspheres was achieved, whose 3D unwrapped form is shown in figure 10(d). Finally, the quantitative phase values of the beads were successfully compared to the values provided by a conventional DHM layout [80], whose 3D unwrapped phase distribution is presented in figure 10(e).

Furthermore, experimental validations for different types of fixed biological samples were performed in Ref. [80]. There, RBC, SS and PC-3 cells were imaged with the previous 10X and 20X microscope lenses, and the clear areas close to the cells were considered as R regions. That forced to consider the spatial multiplexing in different directions for a better implementation of the approach [80].

Finally, an analysis of the coherent noise reduction coming from the use of a partially coherent illumination instead of a coherent one was also carried out in Ref. [80]. For that purpose, the authors compared the standard deviation (STD) values of some background regions present in the retrieved phase maps provided by the SMIM approaches described in Refs. [62] and [80]. The direct comparative was performed for the case of the USAF resolution test target imaged with the 20X/0.46NA objective, achieving a factor of around 10 (STD values from 0.31 rad in Ref. [62] to 0.033 rad in Ref. [80]) in the coherent noise reduction, thus quantifying the better quality reconstruction reached with the use of partially coherent illumination. Additionally, the STD values calculated in Ref. [80] in different clear regions of the phase distributions of the microbeads, RBCs, SSCs and PC-3 cells demonstrated a high phase stability, presenting STD results of about 0.030 rad in most of them, thus reinforcing the results achieved for the case of the USAF test target.

### 3.5 Hilbert- Huang single-shot SMIM

The main drawback of the SMIM approach with partially coherent illumination reported in [80] was the prevention of single-shot capability due to the use of a temporal P-S method for complex amplitude retrieval. That reduced its applicability to static samples, being discarded for the study of dynamic processes such as the analysis of cell migration or particle tracking, just to cite two examples.

However, such disadvantage was overcome in Ref. [81] with the application of the technique named Hilbert-Huang single-shot spatially multiplexed interferometric microscopy (H2S2MIM). H2S2MIM employed a modified version of the Hilbert-Huang phase microscopy (H2PM) technique [92], instead of a temporal P-S approach, for phase map reconstruction from a single slightly off-axis hologram. H2PM is based on the two-dimensional (2D) Hilbert-Huang transform (HHT) [92]. Essentially, 2D HHT is a combination of both empirical mode decomposition (EMD) [93] and 2D Hilbert spiral transform (HST) method [94], a 2D extension of 1D Hilbert transform employed in QPI [94–98]. H2PM utilizes an adaptive image-domain fringe filtering based on 2D enhanced fast EMD [97, 98] (first part of the 2D HHT) and 2D HST with precise phase demodulation aided by the local fringe direction map estimation (second part of the 2D HHT) [98]. Such an algorithm allows efficient analysis for the difficult case of the microbeads (introduce spurious circular diffraction patterns, present significant variation in period and orientation of the fringes, and fringe discontinuities in the borders), by the modifications introduced in [81] to H2PM: 1) employment of the variational image decomposition (VID) technique and Otsu binarization to analyze separately the bead region and the carrier fringe region, 2) the consideration of global parameters such as the mean amplitude and the mean spectral energy to increase the signal-to-noise ratio of the holograms, 3) the use of the fringe direction estimator reported in [92], but additionally employing EMD adaptive filtering of the very noisy first-estimate direction map to obtain both accuracy and smoothness of the automatically estimated fringe direction map, and 4) the use of an iterative phase unwrapping technique based on transport of intensity equation [99] to aid the one proposed in [100].

The experimental implementation was the same as in the previous SMIM approach [80], but with the requirement of just one hologram for S2H2PM application. Nonetheless, the grating was again placed onto a motorized linear translation stage for comparing the results with the provided by the previous temporal P-S algorithm reported in [80]. In that case, a bead of 90 μm in diameter and two different types of PCCs (PC-3 and LnCaP lines) were imaged with an Olympus

UMPlnFl 20X/0.46NA objective. Figure 11 presents the final results for the case of the bead [Row (a)] and the LnCaP cells [Row (b)]. A small ROI of the holograms are included in Column 1, whereas the 3D phase maps provided by H2S2MIM and temporal P-S approach are depicted in Columns 2 and 3, respectively. The quantitative results reached by both methods perfectly matched each other for both cases, leaving aside some background differences, thus validating H2S2MIM for QPI.

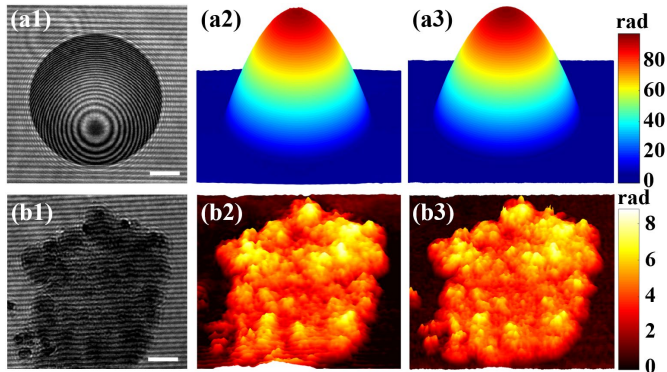

**Figure 11.** Experimental validation of H2S2MIM involving static samples: (a1) and (b1) holograms for the case of a 90 μm sphere and prostate cancer cells, respectively, and (a2)-(b2) and (a3)-(b3) perspective views of unwrapped phase distributions provided by H2S2MIM and by temporal P-S algorithm, respectively. Scale bars represent 20 μm. (Images derived from [81] by permission of OSA).

Finally, single-shot operation principle was demonstrated by performing an experiment involving two microbeads of 45 μm in diameter flowing into a 100 μm thickness counting chamber filled with water. Figure 12 includes the first frames of the full movies (Visualization 1 and Visualization 2) regarding the holograms [figure (12(a)] and the unwrapped phase distributions [figure 12(b)] of a ROI containing the microbeads.

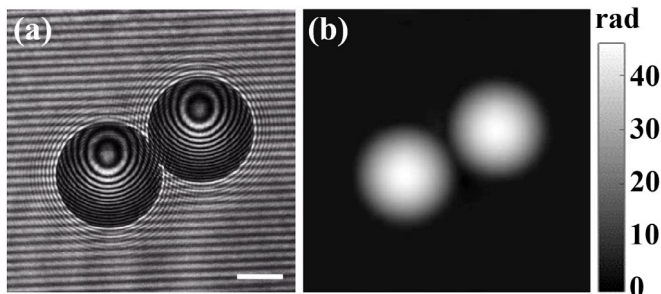

**Figure 12.** Single-shot operation principle demonstration of H2S2MIM involving flowing microbeads: (a) holograms (Visualization 1, 0.5 MB) and (b) unwrapped phase images (Visualization 2, 0.25 MB). Scale bar represents 20 μm. (Movies taken from [81] by permission of OSA).

## 4. Discussion and conclusions

In this contribution, we have presented a review of SMIM technique using a diffraction grating. SMIM is an interferometric technique that updates a regular microscope with holographic capabilities for QPI. SMIM implements a S-M CPI architecture in the embodiment of a standard bright field microscope by introducing three minimal modifications: 1) the replacement of the broadband illumination of the microscope by a coherent or partially coherent light source for allowing interferences; 2) a spatially multiplexed input plane, divided into two or three regions, to transmit in parallel both reference and object beams; and 3) a 1D diffraction grating to provide a holographic recording.

Starting with a demonstrator at the lab of SMIM concept [56, 57, 77], such technique has been implemented in a conventional microscope for different purposes [62, 78–81]: 1) demonstration of the implementation of SMIM concept into a standard microscope [62], 2) resolution enhancement by combining SMIM with SR techniques in the S2MIM technique [78], 3) expanding the range of samples to be inspected with SMIM technique by defining a reflective and transflective imaging mode in the OV-SMIM approach [79], 4) to increase the quality of phase images by reducing the coherent noise with the use of partially coherent illumination and to define a less demanding FOV spatial multiplexing in [80], and 5) to allow single-shot capability of SMIM with partially coherent illumination by means of the application of a phase retrieval method based on HHT in the H2S2MIM approach [81].

The strength of SMIM stems from the CPI implementation in a conventional microscope. When both interferometric beams pass through the same optical system and nearly share the optical path, a low coherence length illumination can be used to reduce the coherent noise. In addition, the instabilities (mechanical and thermal changes) of the system equally affect to both optical paths, thus not disturbing the results. Perhaps the major handicap of SMIM approaches is the requirement of a clear region in the FOV for reference beam transmission. The input plane must be spatially multiplexed into two or three regions (according to the needs), thus making the use of specific chambers desirable for implementation. Currently, the design of customized chambers for specific applications (DHM [59], fluorescence microscopy [101], light-sheet microscopy [102], optofluidic imaging [103], etc.) is quite common and does not represent an additional problem.

In summary, SMIM is a cost-effective, simple, and highly stable way to convert a regular microscope into a holographic one. Such technique has been widely validated involving different resolution test targets, static and dynamic microbeads, and static biosamples. This review has been focused on SMIM approaches using a diffraction grating, but a beam splitter cube can be utilized instead for interferometric recording [82, 83]. SMIM could be very appealing in applications like microfluidic, where the sample flows inside a narrow channel, which can be specifically designed to the SMIM requirements.

Future developments will be focused on the improvement of both S2MIM and OV-SMIM techniques by achieving high-quality imaging and single-shot capability. For that purpose, several partially coherent illuminations will be simultaneously utilized, and an algorithm based on HHT will be employed for phase retrieval. Such progresses could provide, on one hand, superresolved imaging of dynamic processes and, on the other hand, RI and morphology variations in real time. Finally, it is also worth mentioning that SMIM can be combined with other microscopy techniques, such as fluorescence microscopy or conventional bright field imaging for advanced analysis.

## Acknowledgements

The authors want to thank Dr. José Antonio López-Guerrero from the Fundación Instituto Valenciano de Oncología-FIVO for PC-3 biosample preparation, Juan Martínez Carranza and Maria Cywinska for providing phase unwrapping and VID codes, Dr. Alejandra Soriano Portillo from the Instituto de Ciencia Molecular (ICMol) for AFM characterization of the diffraction grating, Prof. Carles Soler and Mr. Paco Blasco from Proiser R+D S.L. for providing the swine sperm sample.

This work was supported by the Spanish Ministerio de Economía y Competitividad and the Fondo Europeo de Desarrollo Regional (FIS2017-89748-P), and in part by the Nationaly Science Center, Poland (OPUS 13 2017/25B/ST7/02049), the Polish National Agency for Academic Exchange and Warsaw University of Technology statutory funds, and by FOTECH-1 project granted by Warsaw University of Technology under the program Excellence Initiative: Research University (ID-UB).

## Ethical statements

The authors declare that there are not conflicts of interest related to this article.

## ORCID iDs

Vicente Micó https://orcid.org/0000-0001-9457-1960
Maciej Trusiak https://orcid.org/0000-0002-5907-0105